\documentclass[letterpaper, 10 pt, conference]{ieeeconf}
\IEEEoverridecommandlockouts 
\usepackage{graphicx} 
\usepackage{cite}
\usepackage{caption}
\usepackage{amsmath}
\usepackage{amsfonts}
\usepackage{dsfont}
\let\labelindent\relax
\usepackage{enumitem,kantlipsum}
\usepackage{subcaption}

\title{\LARGE \bf Scalable Adaptive Traffic Light Control Over a Traffic Network Including Transit Delays}

\author{Yingqing Chen and Christos G. Cassandras
\thanks{Y. Chen and C. G. Cassandras are
with the Division of Systems Engineering and Center for Information and
Systems Engineering, Boston University, Brookline, MA 02446
\tt\small\{yqchenn;cgc\}@bu.edu.}
}

\begin{document}
\maketitle
\thispagestyle{empty}
\pagestyle{empty}

\begin{abstract} 
We study the Traffic Light Control (TLC) problem for a traffic network with multiple intersections in an artery, including the effect of transit delays for vehicles moving from one intersection to the next. The goal is to minimize the overall mean waiting time and improve the ``green wave'' properties in such systems. 
Using a stochastic hybrid system model with parametric traffic light controllers, we use Infinitesimal Perturbation Analysis (IPA) to derive a data-driven cost gradient estimator with respect to these parameters. We then iteratively adjust them through an online gradient-based algorithm. We show that the event-driven nature of the IPA estimators driving the controllers leads to scalable computationally efficient controllers as the dimensionality of the traffic network increases. 
\end{abstract}

\section{Introduction}
The Traffic Light Control (TLC) problem entails dynamically adjusting the traffic light cycles in an intersection or a set of intersections in order to improve the overall traffic system performance (normally measured through one or more congestion metrics). The single intersection TLC has been thoroughly studied using different approaches such as model-based optimization \cite{zhang_traffic_2017},\cite{van2007integrated}, computational intelligence \cite{kaur_adaptive_2014},\cite{chu_multi-agent_2020}, and online optimization methods \cite{fleck_adaptive_2016}. However, the transition from one to multiple interconnected intersections is particularly challenging for at least three reasons: $(i)$ Extensions from one-dimensional to high-dimensional TLC problem solutions which are \emph{scalable} are not easy to obtain, $(ii)$ When the distance between adjacent intersections is relatively short, traffic is often blocked and such blocking effects must be taken into account, and $(iii)$ The transit delays experienced by traffic between intersections must be accounted for in order to effectively coordinate an intersection with its downstream counterpart(s). 

Starting with simple multi-intersection scenarios, \cite{gershenson_self-organizing_2005} presents three traffic-responsive methods by which traffic lights are self-organized, and \cite{hassin_flow_1996} abstracts the fixed cycle, fixed pattern multi-intersection traffic signal problem as a network synchronization problem, assuming the cost only depends on the pairwise cycle offset of adjacent traffic lights. Adding more controllable parameters, several approaches have developed centralized controllers
emphasizing the synchronization effect over multiple traffic lights, e.g., a steady-state signal control approach in \cite{he_steady-state_2015} and an adaptive linear quadratic regulator in \cite{wang_optimizing_2022}.
To deal with the high computational intensity of such centralized optimization methods, 
various decomposition methods have been proposed \cite{lin_efficient_2012},\cite{zhou_multiagent_2015},\cite{huang_adaptive_2018}.
To address the computational complexity caused by non-linearities, \cite{le_linear-quadratic_2013} presents a linear prediction model, and expresses the MPC controller using a QP formulation, which quadratically penalizes the number of vehicles in the network and linearly penalizes the control decision.

Although many methods have been implemented to reduce computational requirements, especially in large traffic networks, \cite{BazzanAnaL.C.2005ADAf} argues that centralized approaches to traffic signal control cannot cope with the increasing complexity of urban traffic networks. Thus, decentralized approaches have also been investigated, including multi-agent reinforcement learning approaches \cite{wang_stmarl_2022},\cite{el-tantawy_multiagent_2013}.

The model-based optimization methods mentioned above must deal with the ``curse of dimentionality'' due to the nature of a centralized approach, while in learning-based methods, the computational load is shifted to the training of controllers, which also requires a large amount of historical data and processing, especially for large-scale urban traffic networks.
With this motivation, the first contribution of this paper is to exploit the scalability properties of the single-intersection adaptive data-driven TLC approach in \cite{fleck_adaptive_2016},\cite{chen2023adaptive} based on Infinitesimal Perturbation Analysis (IPA) used to estimate performance gradients with respect to the parameters of a TLC controller.
These gradient estimates are then used to iteratively seek optimal values for GREEN cycles at each intersection. 
In particular, since IPA-based gradient estimators are entirely \emph{event-driven}, these algorithms scale with the (relatively small) number of events in each system intersection, not the (much larger) state space dimensionality \cite{chen_stochastic_2020}\cite{chen2023adaptive}. Moreover, IPA is independent of any modeling assumptions regarding the stochastic processes characterizing traffic demand and vehicle behavior, driven only by actual observed traffic data similar to learning-based approaches.
The second contribution of the paper is to incorporate vehicle transit delays into the traffic network modeled as a stochastic hybrid system where the controllable traffic light switching process is event-driven, while the dynamics of the vehicle flows are time-driven. Similar to the model in \cite{chen_stochastic_2020}, the traffic flow at each intersection depends on the incoming flow from an upstream intersection delayed by the transit time between intersections. This delayed flow joining process was modeled in \cite{chen_stochastic_2020} through multiple stages of ``servers''. In this paper, we design a simpler framework that can capture the flow delay process while reducing the number of events required. This allows us to take advantage of one of the fundamental properties of IPA \cite{cassandras_perturbation_2010} applied to network systems, whereby the effect of a parameter perturbation at one traffic light can only propagate to adjacent traffic lights as a result of a limited and easily tractable number of events; this facilitates the estimation of network-wide performance gradients with respect to a parameter at a specific traffic light.

The remainder of this paper is organized as follows. In section II, we formulate the TLC problem for a serial multi-intersection network and present the stochastic hybrid system modeling framework. Section III details the derivation of the IPA estimators for a network-wide cost function gradient with respect to a controllable parameter vector. The IPA estimators are then incorporated into a gradient-based optimization algorithm. In Section IV, we conduct multiple simulation experiments under different settings and demonstrate the adaptivity and scalability properties of this approach. Finally, we conclude and discuss future work in Section V.

\section{Problem Formulation}\label{sec:Problem Formulation}
\begin{figure}
    \centering
    \includegraphics[width=0.9\columnwidth]{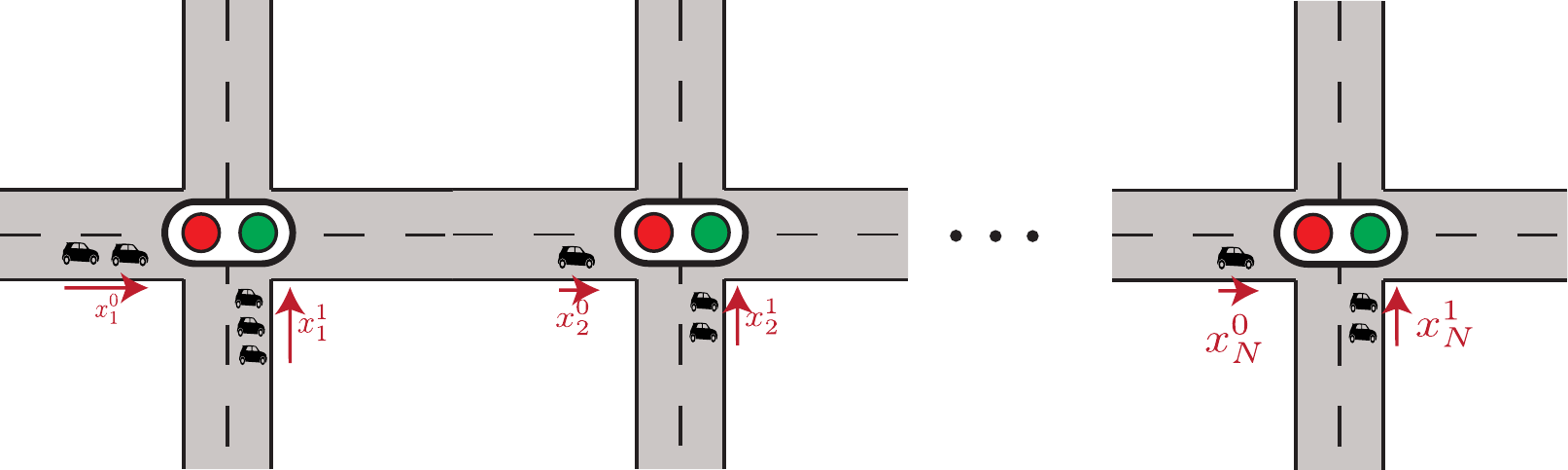}
    \caption{Traffic system with $N$ intersections}
    \label{fig:intersections}
\end{figure}

We consider $N$ consecutive intersections in an artery as shown in Fig.\ref{fig:intersections}. In this paper, we limit our analysis to a model with the following simplifications: left-turn and right-turn traffic flows are not considered; the YELLOW cycle is combined with the RED one; the road between any two intersections is long enough such that no blocking occurs; for each intersection $n$ ($n=1,\ldots,N$), we consider only two vehicle flows (perpendicular to each other) indexed by direction $d=0, 1$, where $d=0$ represents the artery direction (West-East) and $d=1$ represents the side road direction (South-North). Each road segment is modeled as a queue where vehicles stop when facing a RED light. Thus, we define a state vector for intersection $n=1,\ldots,N$: $x_n(t)=[x_n^0(t), x_n^1(t)]$, where $x_n^d(t) \in \mathbb{R}_0^+$, ($d=0,1$) denotes the flow content of the queue for direction $d$ at intersection $n$.

We model the input to each queue as a flow process
$\{ \alpha_n^d(t) \}$ where $\alpha_n^d(t)$ is the \emph{stochastic} instantaneous arrival rate of direction $d$ at intersection $n$, independent of all queue states. Similarly, we denote the departure flow process as $\{ \beta_n^d(t) \}$. The departure process is dependent on the corresponding traffic light control denoted by $u_n^d(t)$, where $u_n^d(t)=1$ denotes a GREEN phase for direction $d$ at intersection $n$, and $u_n^d(t)=0$ denotes RED phase accordingly. Note that the controller is designed to always satisfy $u_n^0(t)+u_n^1(t)=1$. We also define $u_n^d(t)$ to be right-continuous in order to accurately represent the control policy defined in the sequel. Thus, $\beta_n^d(t)$ can be expressed as follows:
\begin{align} \label{beta}
  \beta_n^d(t) =
    \begin{cases}
      h_n^d(t), & \text{if $x_n^d(t)>0$ and $u_n^d(t)=1$}\\
      \alpha_n^d(t), & \text{if $x_n^d(t)=0$ and $u_n^d(t)=1$}\\
      0, & \text{otherwise}
    \end{cases}   
\end{align}
where $n=1,\ldots,N, d=0,1$, and $h_n^d(t)$ is the unconstrained departure rate for which appropriate models can be used (see Section IV).
When considering a single intersection, \{$\alpha_n^d$\} is an exogenous arrival process, i.e., assumed independent of the state of the system. In the multi-intersection case, this is only valid for $n=1,\ldots,N,d=1$ (all side-road directions) or for $n=1$, $d=0,1$ (both directions at the first intersection). Otherwise, the arrival flow process is determined by the departure process of the upstream intersection. In Fig. \ref{fig:intersections}, for example, $\alpha_2^0(t)$ is the instantaneous rate of the artery flow departing intersection $n=1$ joining the tail of the artery queue at intersection $n=2$ at time $t$. 
We can then write the queue flow dynamics as
\begin{equation}
\label{eqn:xdynamic}
  \dot x_n^d(t) =\alpha_n^d(t)-\beta_n^d(t), ~~~~~n=1,\ldots,N, ~d=0,1
\end{equation}
where $\alpha_n^0(t)$ for $n>1$ will be defined in the sequel so as to properly capture its dependence on $\beta_{n-1}^0(t)$ and included the effect of flow transit delays.

Moreover, we define clock state variables $z_n^d(t)$, $d=0,1$, $n=1,\ldots,N$ to measure the time since the last switch from RED to GREEN for each corresponding vehicle flow. To be consistent with $u_n^0(t)+u_n^1(t)=1$, we require that 
$z_n^{(1-d)} (t)>0$ when $z_n^d(t)=0$ for all $t\ge0$. The dynamics of $z_n^d(t)$ are:
\begin{equation}
\label{eqn:zdynmic}
  \dot z_n^d(t) =
    \begin{cases}
      1, & \text{if $u_n^d(t)=1$ }\\
      0, & \text{otherwise} 
    \end{cases}       
\end{equation}
Note that $\dot z_n^0(t) + \dot z_n^1(t)=1$ always holds. In addition, we define $z_n^d(t)$ to be left-continuous, so that when the light switches from GREEN to RED, $z_n^d(t)>0$,  $z_n^d(t^+)=0$, and $u_n^d(t)=0$ (since $u_n^d(t)$ is right-continuous). 

\textbf{Traffic Light Control.} We define a threshold parameter $\theta_n^d > 0$ for each direction $d$ of each intersection $n$ indicating current GREEN cycle time of each direction, and an associated vector:
\begin{equation} \label{parametervector}
    \Theta = [\theta_1^0, \theta_1^1, \theta_2^0, \theta_2^1,\ldots,\theta_N^0, \theta_N^1]
\end{equation}
The traffic light controller $u_n^d(t)$ switches from GREEN to RED when $z_n^d(t)$ reaches the threshold $\theta_n^d$:
\begin{equation}
\label{eqn:control rule}
  u_n^d(t) =
    \begin{cases}
      1, & \text{if [$z_n^d(t)\in (0, \theta_n^d)$ and $z_n^{1-d}(t)=0$]}\\
      & \text{ OR [$z_n^{1-d}=\theta_n^{1-d}$]} \\
      0, & \text{otherwise}
    \end{cases}       
\end{equation}
In prior work for a single intersection (\cite{fleck_adaptive_2016},\cite{chen2023adaptive}), $\Theta$ includes additional controllable parameters that give the controller quasi-dynamic properties that improve its performance. We omit these in this paper so as to focus on the multi-intersection aspects of the controllers, but note they are straight-forward to include in the next steps of this research. 

\textbf{Events.} The state transitions in the model defined through (\ref{beta}), (\ref{eqn:xdynamic}), and (\ref{eqn:zdynmic}) under the controller (\ref{eqn:control rule}) are dictated by several events defined as follows (see also Table \ref{table:events}). 

\emph{Basic events.} For $n=1,\ldots,N, d=0,1$: (a) [$x_n^d\downarrow 0$]: $x_n^d(t)$ reaches 0 from above,
(b) [$x_n^d\uparrow 0$]: $x_n^d(t)$ becomes positive from 0,
(c) [$z_n^d\uparrow \theta_n^d$]: $z_n^d$ reaches its upper bound,
(d) [$\alpha_n^d\uparrow 0$]: $\alpha_n^d$ becomes positive from 0,
(e) [$\alpha_n^d\downarrow 0$]: $\alpha_n^d$ reaches 0 from above.

\emph{Light switching events. } It is convenient to define the following light switching events which are induced by  [$z_n^d\uparrow \theta_n^d$] events:
\begin{itemize}
    \item $G2R_n^d$: traffic light for direction $d$ at intersection $n$ switches from GREEN to RED. This is triggered by [$z_n^d\uparrow \theta_n^d$], $n=1,\ldots,N$, $d=0,1$. 
    \item $R2G_n^d$: traffic light for direction $d$ at intersection $n$ switches from RED to GREEN. This is triggered by [$z_n^{1-d}\uparrow \theta_n^{1-d}$], $n=1,\ldots,N$, $d=0,1$. 
\end{itemize}

\subsection{Flow Burst Modeling and Analysis}
A major consideration in studying a multi-intersection system is the role of a \emph{flow burst} which is generated at each intersection and has impact on the downstream intersection. This was first studied in \cite{chen_stochastic_2020} where a complicated sequence of sub-processes was used limiting its use to only two intersections. Here, we provide a more direct way to model such flow bursts that easily extends to multiple intersections.
Specifically, when a RED light switches to GREEN at $n$, a new flow burst is generated consisting of all vehicles queued at $n$ that are released. We regard this as a single flow with continuously changing rate. Therefore, the impact this upstream flow has on the downstream queue $n+1$ is caused by the start and end of this flow generation at each GREEN phase. 

We start by defining event $G_{n,m}$, $n=1,\ldots,N-1$, $m=1,2,\ldots$ causing the generation of an artery flow burst from intersection $n$ during the $m$th GREEN cycle. This event is induced at time $t$ by one of two events: (a) $R2G_n^0$ if $x_n^0(t)>0$, or (b) the first [$\alpha_n^0\uparrow 0$] event within the $m$th GREEN cycle if $x_n^0(t)=0$. Note that $G_{n,m}$ only affects the artery's intersections, hence the $d$ indicator is omitted.

When $G_{n,m}$ occurs, $\beta_n^0(t)$ increases from zero to positive and the corresponding flow burst will join the downstream queue $n+1$ (empty or not) following a time delay, which then results in the downstream arrival rate increasing from 0 to positive, thus creating an interdependence of flow rates between neighboring intersections. 
The delay depends on (a) the average speed of the flow burst, denoted by $v_n$, affected by road quality and driver behavior,
(b) the road length $L_n$ between intersections $n$ and $n+1$, and 
(c) the average length $l$ of a vehicle (including a safe distance between two vehicles).
Then, the flow process relationship between any two adjacent intersections caused by such delay can be estimated as:
\begin{equation}\label{eqn:flow_process_relationship}
    \alpha_{n+1}^0(t) = \beta_n^0(t-\frac{L_n-x_{n+1}^0(t)*l}{v_n})
\end{equation}
for $n=1,\ldots,N-1$. Due to the no blocking assumption, $L_n-x_{n+1}^0(t)*l>0$ always holds, which implies that $\alpha_{n+1}^0(t)$ is always dependent on the upstream output process. 
For convenience, set $\Delta_n(t)=\frac{L_n-x_{n+1}^0(t)*l}{v_n}$ to be the delay in
(\ref{eqn:flow_process_relationship})
and note that $\Delta_n(t)\in (0,\frac{L_n}{v_n}]$. 
We also assume that any two flow bursts generated from the same intersection but different GREEN cycles will not join each other before the front one joins the downstream queue. 

\begin{figure}
    \centering
    \includegraphics[width=0.8\columnwidth]{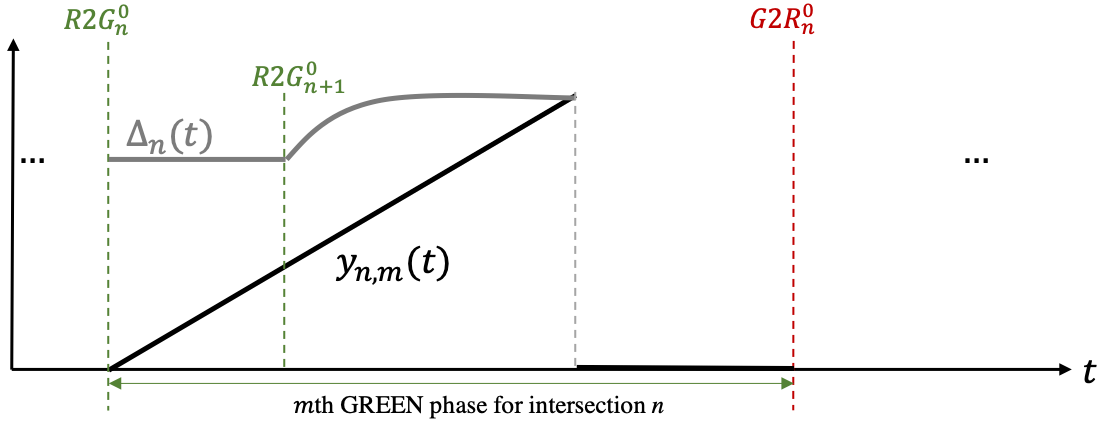}
    \caption{Example for trajectories of $y_{n,m}(t)$ and $\Delta_n(t)$ }
    \label{fig:y}
\end{figure}

We now define a second clock variable $y_{n,m}(t)$, similar to (\ref{eqn:zdynmic}), which denotes the time elapsed since $G_{n,m}$ occurs. Its dynamics are
\begin{equation}\label{eqn:ydynmic}
  \dot y_{n,m}(t) =
    \begin{cases}
      1, & \text{if $y_{n,m}(t)\in (0,\Delta_n(t))$}\\
      0, & \text{otherwise}\\
    \end{cases}  
\end{equation}
Unlike $z_n^d(t)$ in (\ref{eqn:zdynmic}), $y_{n,m}(t)$ has a \emph{single} cycle for each flow burst $m$. When $G_{n,m}$ occurs,
we set $y_{n,m}(t)=0$ and initialize (\ref{eqn:ydynmic}) so that $y_{n,m}(t^+)>0$. 
As soon as $y_{n,m}(t)=\Delta_n(t)$, we set $y_{n,m}(t^+)=0$. An example is shown in Fig.\ref{fig:y}. 
Note that $y_{n,m}(t)>0$ is possible for multiple different $m$, i.e., several flow bursts may be active generated from the same intersection but during different GREEN cycles. This may happen when $L_n$ is large and the flow bursts have yet to reach the downstream queue.
Also note that when $y_{n,m}(t)=\Delta_n(t)$, this implies that the head of a flow burst generated by event $G_{n,m}$ joins the downstream artery queue $n+1$ at time $t$. This is the first instant when this
flow burst can have an impact on the downstream queue. Therefore, we define it as an event $J_{n,m}$, $n=1,\ldots,N-1$, $m=1,2,\ldots$, which also induces an event [$\alpha_{n+1}^d\uparrow 0$], i.e., the input flow at $n+1$ becomes positive again.

Similarly, in order to model the end of a flow burst, we define an event $G_{n,m}^e$, $n=1,\ldots,N-1$, $m=1,2,\ldots$ which indicates the end of an artery flow burst generation from intersection $n$ during the $m$th GREEN cycle. This can be induced by (a) $G2R_n^0$ if either $\alpha_n^0(t)>0$ or $x_n^0(t)>0$, or (b) the last [$\alpha_n^0\downarrow0$] event inside the $m$th GREEN cycle if $x_n^0(t)=0$. 
This leads to the definition of another clock state variable $r_{n,m}(t)$ which measures the time elapsed since $G_{n,m}^e$ occurs. Its dynamics are:
\begin{equation}\label{eqn:rdynmic}
  \dot r_{n,m}(t) =
    \begin{cases}
      1, & \text{if $r_{n,m}(t)\in(0,\Delta_n(t))$}\\
      0, & \text{otherwise}\\
    \end{cases}  
\end{equation}
When $G_{n,m}^e$ occurs, we set $r_{n,m}(t)=0$ and initialize (\ref{eqn:rdynmic}) so that
$r_{n,m}(t^+)>0$. As soon as $r_{n,m}(t)=\Delta_n(t)$, we set $r_{n,m}(t^+)=0$. 
Similar to $y_{n,m}(t)$, it is possible that $r_{n,m}(t)>0$ for multiple different $m$.
Note that when $r_{n,m}(t)=\Delta_n(t)$, this implies that the flow burst generated from the $m$th GREEN cycle at intersection $n$ ceases to have any impact on the downstream queue. Therefore, we define it as an event $J_{n,m}^e$,
$n=1,\ldots,N-1$, $m=1,2,\ldots$, which also induces an event [$\alpha_{n+1}^d \downarrow 0$].
Since we consider a single flow burst generated by the same GREEN phase, any [$\alpha_{n+1}^0\uparrow0$] or [$\alpha_{n+1}^0\downarrow0$] between $J_{n,m}$ and $J_{n,m}^e$ is exogenous and has no impact on the downstream queue state.

In summary, the generation and impact of a flow burst is captured through the four additional \emph{flow burst tracing events} $G_{n,m}$, $J_{n,m}$, $G_{n,m}^e$, $J_{n,m}^e$, which facilitate the IPA gradient evaluation in Section \ref{section:ipa}. The full list of events is shown in Table \ref{table:events}. 
We now have a state vector [$x_n(t), z_n(t), y_n(t), r_n(t)$] for intersection $n$ where
$z_n(t) = [z_n^0(t), z_n^1(t)]$, $z_n^d(t) \in \mathbb{R}_0^+$ for $d=0,1$,
$y_n(t)=[y_{n,1}(t), y_{n,2}(t),\ldots]$, $y_{n,m}(t) \in \mathbb{R}_0^+$ for $m\in \mathbb{N}^{+}$,
$r_n(t)=[r_{n,1}(t), r_{n,2}(t),\ldots]$, $r_{n,m}(t) \in \mathbb{R}_0^+$ for $m\in \mathbb{N}^{+}$.
Therefore, we define the traffic light controller for intersection $n$ as:
\begin{align} \label{udef}
    u_n(x_n(t), z_n(t), y_n(t), r_n(t)) = [u_n^0(t), u_n^1(t)]
\end{align}
where $u_n^d(t)$ denotes the control for direction $d$ at intersection $n$. In particular, $u_n^d(t)=1$ denotes a GREEN phase, and $u_n^d(t)=0$ denotes a RED phase accordingly.

\begin{table}[]
\centering
\caption{List of Events}
\resizebox{0.5\textwidth}{!}{%
\begin{tabular}{|l|l|}
\hline
Basic Events & [$x_n^d\downarrow 0$], [$x_n^d\uparrow 0$], [$z_n^d\uparrow \theta_n^d$],[$\alpha_n^d\uparrow 0$], [$\alpha_n^d\downarrow 0$] \\ \hline
    Light Switching Events & $R2G_n^d$, $G2R_n^d$ \\ \hline
    Flow Burst Tracing Events & $G_{n,m}$, $J_{n,m}$, $G_{n,m}^e$, $J_{n,m}^e$ \\ \hline
\end{tabular}%
}
\label{table:events}
\end{table}


The multi traffic light intersection system can be viewed as a hybrid system in which the time-driven dynamics are
given by (\ref{eqn:xdynamic}), (\ref{eqn:zdynmic}), (\ref{eqn:ydynmic}), (\ref{eqn:rdynmic}) and (\ref{beta}), while the event-driven dynamics are dictated by the basic events in Table \ref{table:events}; these induce
associated light switching and flow burst tracing events (defined for convenience). Although the dynamics are based on knowledge of the instantaneous flow processes $\{ \alpha_n^d(t) \}$ and $\{ \beta_n^d(t) \}$, we will show that the IPA-based adaptive controller we design does not require such knowledge and depends only on estimating some rates in the vicinity of certain critical observable events.


\begin{figure}
    \centering
    \includegraphics[width=0.8\columnwidth]{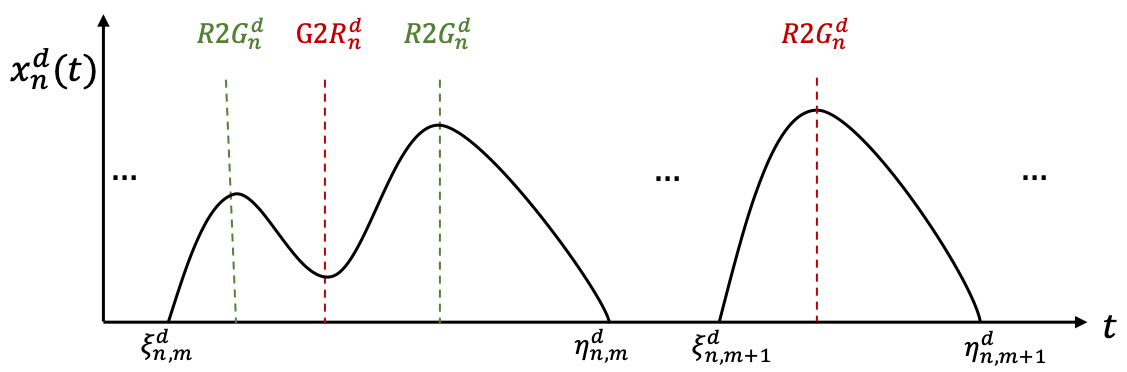}
    \caption{Typical sample path of a traffic queue}
    \label{fig:sample path}
\end{figure}

\subsection{TLC Optimization Problem}
With the parameterized controller defined above, our aim is to optimize a performance metric for the intersection operation with respect to these controllable parameters that comprise the vector $\Theta$ defined in (\ref{parametervector}). 
We choose our performance metric to be the weighted mean of all queue lengths over a fixed time interval $[0,T]$:
\begin{equation} \label{Lfunction}
    L(\Theta;  x(0),z(0), T) = \frac{1}{T}\sum_{n=1}^{N}\sum_{d=0}^{1}\int_{0}^{T} \omega_n^d x_n^d(\Theta,t) \,dt 
\end{equation}
where $\omega_n^d$ is a weight associated with a direction $d$ queue at intersection $n$.  In order to focus on the structure of a typical sample path of the hybrid system, observe that the sample path of any flow queue content $\{x_n^d(t)\}$ consists of alternating \emph{Non-empty Periods} (NEPs) and \emph{Empty Periods} (EPs), which correspond to time intervals when $x_n^d(t)>0$ and $x_n^d(t)=0$ respectively, as shown in Fig. \ref{fig:sample path}. We define two additional events: $S_n^d$ for starting NEPs 
and $E_n^d$ for ending them, both induced by basic events defined earlier. 
Moreover, we denote the $k$th NEP of queue $(n,d)$ by $[\xi_{n,k}^d, \eta_{n,k}^d)$ where $\xi_{n,k}^d$, $\eta_{n,k}^d$ are the occurrence times of the $k$th $S_n^d$ event and $k$th $E_n^d$ event respectively. 
Since $x_n^d(t)=0$ during EPs of queue $(n,d)$, the sample function $L(\Theta;  x(0),z(0), T)$ in (\ref{Lfunction}) can be rewritten as
\begin{equation}
\label{eqn:L}
    L(\Theta; x(0),z(0), T) = \frac{1}{T}\sum_{n=1}^{N}\sum_{d=0}^{1}\sum_{k=1}^{K_n^d}\int_{\xi_{n,k}^d}^{\eta_{n,k}^d} \omega_n^d x_n^d(\Theta,t) \,dt 
\end{equation}
where $K_n^d$ is the (random) total number of NEPs during the sample path of queue $(n,d)$ over $[0,T]$.

Thus, our goal is to determine $\Theta$ that minimizes the expected weighted mean queue length:
\begin{equation}
    J(\Theta; x(0),z(0), T) = E[L(\Theta;  x(0),z(0), T)]
\end{equation}
We note that it is not possible to derive a closed-form expression of $J(\Theta; x(0), z(0), T)$ even if we had full knowledge of the processes $\{ \alpha_n^d(t) \}$ and $\{ \beta_n^d(t) \}$. Therefore, a closed-form expression for the gradient $\nabla J(\Theta)$ is also infeasible. 
The role of IPA is to obtain an \emph{unbiased} estimate of $\nabla J(\Theta)$ based on the sample function gradient $\nabla L(\Theta)$ which can be evaluated based only on data directly observable along a single sample path such as Fig. \ref{fig:sample path}, as will be shown in the next section. The unbiasedness of $\nabla L(\Theta)$ is ensured
under mild conditions on $L(\Theta)$ (see \cite{cassandras_perturbation_2010}) and assuming that $\alpha_n^d(t)$ are piecewise continuously differentiable in $t$ w.p. 1. In particular, we emphasize that no explicit knowledge of $\alpha_n^d(t)$ is necessary to estimate $\nabla J(\Theta)$ through $\nabla L(\Theta)$.

We can now invoke a gradient-based algorithm of the form
\begin{equation} \label{gradientopt}
    \Theta_{i,l+1} = \Theta_{i,l}-\rho_l \big[\frac{dJ}{d\Theta_{i,l}}\big]_{IPA}
\end{equation}
where $\Theta_{i,l}$ is the $i$th parameter of $\Theta$ at the $l$th iteration ($i=1,\ldots,2N$), $\rho_l$ is the stepsize at the $l$th iteration, and $[\frac{dJ}{d\Theta_{i,l}}]_{IPA}$ is the IPA estimator of $\frac{dJ}{d\Theta_{i,l}}$, which will be derived in the next section.

\section{Infinitesimal Perturbation Analysis} \label{section:ipa}

We begin with a brief review of the IPA framework in \cite{cassandras_perturbation_2010}. Consider a sample path over $[0,T]$ and denote the occurrence time of the $k$th event (of any type) by $\tau_k$. Let $x'(\theta, t)$, $\tau'_k(\theta
)$ be the derivatives of $x(\theta, t)$, $\tau_k(\theta)$ over the scalar controllable parameter of interest $\theta$ respectively. We omit the dependence on $\theta$ for ease of notation hereafter. The dynamics of $x(t)$ are fixed over any inter-event interval$ [\tau_k, \tau_{k+1})$, represented by $\dot{x}(t)=f_k(t)$. Then, the state derivative satisfies 
\begin{equation}
\label{eqn:state_derivative_pre}
    \frac{d}{dt}x'(t)=\frac{\partial f_k(t)}{\partial x}x'(t) + \frac{\partial f_k(t)}{\partial \theta}
\end{equation}
with boundary condition:
\begin{equation}
\label{eqn:state_derivative}
    x'(\tau_k^+) = x'(\tau_k^-) + [f_{k-1}(\tau_k^-)-f_k(\tau_k^+)]\tau_k'
\end{equation}
In order to evaluate (\ref{eqn:state_derivative}), $\tau_k'$ must be determined, which depends on the type of event taking place at $\tau_k$.
For exogenous events (events causing a discrete state transition that is independent of any controllable parameter), we have $\tau_k'=0$.
For endogenous events (events that occur when there exists a continuously differentiable function $g_k$ such that $\tau_k=min\{t>\tau_{k-1}:g_k(x(\theta, t),\theta)=0\}$) with guard condition $g_k = 0$(see \cite{cassandras_perturbation_2010}):
\begin{equation}
\label{eqn:event_time_derivative}
    \tau_k' = -[\frac{\partial g_k}{\partial x} f_k(\tau_k^-)]^{-1}(\frac{\partial g_k}{\partial \theta}+\frac{\partial g_k}{\partial x}x'(\tau_k^-))
\end{equation}
This framework captures how system states and event times change with respect to controllable parameters. Our goal is to estimate $\nabla J(\Theta)$ through $\nabla L(\Theta)$, and, according to (\ref{eqn:L}), the performance metric expression is a function of event time and system state variables. Thus, we apply the IPA framework to the TLC problem and evaluate how a perturbation in $\theta$ would affect performance metrics.

\subsection{State Derivatives.}\label{sec:state derivative}
We define the derivatives of the state $x_n^d(t)$, $z_n^d(t)$, $y_{n,m}(t)$, $r_{n,m}(t)$ and event time $\tau_k$ with respect to parameter $\Theta_i$ ($i=1,\ldots,2N$) as follows:
\begin{multline}
     {x_{n,i}^d}'\equiv \frac{\partial x_n^d(t)}{\partial \Theta_i},~ 
     {z_{n,i}^d}'\equiv \frac{\partial z_n^d(t)}{\partial \Theta_i},~
    y_{n,m,i}'\equiv \frac{\partial y_{n,m}(t)}{\partial \Theta_i}, \\
    r_{n,m,i}'\equiv \frac{\partial r_{n,m}(t)}{\partial \Theta_i},~~ 
   \tau_{k,i}'\equiv \frac{\partial \tau_k}{\partial \Theta_i} 
\end{multline}
For ease of notation, we denote the state dynamics in (\ref{eqn:xdynamic}), (\ref{eqn:zdynmic}), (\ref{eqn:ydynmic}), (\ref{eqn:rdynmic}) over an inter-event interval $t\in [\tau_k, \tau_{k+1})$ as follows:
\begin{multline}
    \dot x_n^d(t) = f_{n,d,k}^x(t),~ \dot z_n^d(t) = f_{n,d,k}^z(t),\\ \dot y_{n,m}(t) = f_{n,m,k}^y(t),~  \dot r_{n,m}(t) = f_{n,m,k}^r(t), n=1,2,..N
\end{multline}
Combining the dynamics in (\ref{eqn:xdynamic}), (\ref{eqn:zdynmic}), (\ref{eqn:ydynmic}) and (\ref{eqn:rdynmic}) with (\ref{eqn:state_derivative_pre}), similar to the analysis in \cite{fleck_adaptive_2016} we can easily conclude that the state derivative of any queue is unaffected within any inter-event time interval, i.e., for $t\in[\tau_k,\tau_{k+1})$:
\begin{multline}\label{eqn:derivative within mode}
    {x_{n,i}^d}'(t)={x_{n,i}^d}'(\tau_k^+),\  
    {z_{n,i}^d}'(t)={z_{n,i}^d}'(\tau_k^+),\\ 
    y'_{n,m,i}(t)=y'_{n,m,i}(\tau_k^+),\  
    r'_{n,m,i}(t)=r'_{n,m,i}(\tau_k^+)
\end{multline}
Next, for any discrete event time $\tau_k$, we evaluate queue content derivatives for any possible event occurring to start/end an EP/NEP or within any EP/NEP, and for any controllable parameter $\Theta_i$ ($i=1,\ldots,2N$):
\begin{itemize}
    \item [1)] Event inside EP: Since $x_n(t)=0$ throughout the whole EP, it immediately follows that
    \begin{equation}\label{eqn:state_der_inside_EP}
        {x_{n,i}^d}'(\tau_k^+)=0
    \end{equation}
    
    \item[2)] Event starting EP ($E_n^d$): This is induced by [$x_n^d\downarrow 0$]. The state dynamics change from $f_{n,d,k-1}^x(\tau_k^-)=\alpha_n^d(\tau_k)-h_n^d(t)$ to $f_{n,d,k}^x(\tau_k^+)=\alpha_n^d(\tau_k)-\alpha_n^d(\tau_k)=0$.
    Then, from (\ref{eqn:state_derivative}),
    \begin{equation}
    \label{eqn:state_der_En}
        {x_{n,i}^d}'(\tau_k^+)={x_{n,i}^d}'(\tau_k^-)+(\alpha_n^d(\tau_k)-h_n^d(\tau_k))\tau_{k,i}'
    \end{equation}
    
    \item[3)] Event starting NEP ($S_n^d$): This can be induced in three possible ways:
    \begin{itemize}
        \item [3.1)]$S_n^d$ induced by light switching to RED($G2R_n^d$) when $\alpha_n^d(\tau_k)>0$. We have $f_{n,d,k-1}^x(\tau_k^-)=0$ and $f_{n,d,k}^x(\tau_k^+)=\alpha_n^d(\tau_k)$. Based on (\ref{eqn:state_der_inside_EP}) and (\ref{eqn:state_derivative}):
        \begin{equation}\label{eqn:state_der_Sn_light_switch}
            {x_{n,i}^d}'(\tau_k^+)=-\alpha_n^d(\tau_k)\tau_{k,i}' 
        \end{equation}
        
        \item [3.2)]$S_n^d$ induced by $J_{n-1,m}$ when $d=0, n\ge2$. The state dynamics are $f_{n,d,k-1}^x(\tau_k^-)=0$, $f_{n,d,k}^x(\tau_k^+)=\alpha_n^d(\tau_k^+)-\beta_n^d(\tau_k)$, where $\beta_n^d(\tau_k)$ is defined in (\ref{beta}) and $\alpha_n^d(\tau_k^+)-\beta_n^d(\tau_k)>0$ in order to induce $S_n^d$. Based on (\ref{eqn:state_derivative}) we get
        \begin{equation}\label{eqn:state_der_Sn_alpha}
            {x_{n,i}^d}'(\tau_k^+)= (\beta_n^d(\tau_k)-\alpha_n^d(\tau_k^+))\tau_{k,i}'
        \end{equation}
        
        \item [3.3)] $S_n^d$ induced by an exogenous change in $\alpha_n^d(\tau_k)$ when $n=1,\ldots,N$, $d=1$ or $n=1, d=0,1$. In this case, $\tau_{k,i}'=0$, so that we have
            \begin{equation}
        \label{eqn:state_der_Sn_exo}
            {x_{n,i}^d}'(\tau_k^+) = {x_{n,i}^d}'(\tau_k^-)=0
        \end{equation}
    \end{itemize}

    \item[4)] Event inside NEP. The following are all possible cases:
    \begin{itemize}
        \item [4.1)] $G2R_n^d$ when $n=1,\ldots,N$, $d=0,1$: the state dynamics are $f_{n,d,k-1}^x(\tau_k^-)=\alpha_n^d(\tau_k)-h_n^d(\tau_k)$ and $f_{n,d,k}^x(\tau_k^+)=\alpha_n^d(\tau_k)$. Therefore,
        \begin{equation}\label{eqn:state_der_NEP_G2R}
            {x_{n,i}^d}'(\tau_k^+)={x_{n,i}^d}'(\tau_k^-)-h_n^d(\tau_k)\tau'_{k,i}
        \end{equation}
            
        \item[4.2)]$R2G_n^d$ when $n=1,\ldots,N$, $d=0,1$: the state dynamics are $f_{n,d,k-1}^x(\tau_k^-)=\alpha_n^d(\tau_k)$ and $f_{n,d,k}^x(\tau_k^+)=\alpha_n^d(\tau_k)-h_n^d$. Therefore,
      \begin{equation}\label{eqn:state_der_NEP_R2G}
            {x_{n,i}^d}'(\tau_k^+)={x_{n,i}^d}'(\tau_k^-)+h_n^d(\tau_k)\tau'_{k,i}
        \end{equation}
    
        \item [4.3)]$J_{n-1,m}$ when $n\ge2$, $d=0$: the state dynamics are: $f_{n,d,k-1}^x(\tau_k^-)=-\beta_{n}^d(\tau_k)$, $f_{n,d,k}^x(\tau_k^+)=\alpha_n^d(\tau_k^+)-\beta_n^d(\tau_k)$ where $\beta_n^d(\tau_k)$ follows (\ref{beta}). So that
        \begin{equation}\label{eqn:state_der_NEP_J0}
            {x_{n,i}^d}'(\tau_k^+)=
                {x_{n,i}^d}'(\tau_k^-)-\alpha_{n}^d(\tau_k^+)\tau_{k,i}'
        \end{equation}
    
        \item [4.4)] $J_{n-1,m}^e$ when $n\ge2$, $d=0$: similarly, the state dynamics are: $f_{n,d,k-1}^x(\tau_k^-)=\alpha_{n}^d(\tau_k^-)-\beta_{n}^d(\tau_k)$, $f_{n,d,k}^x(\tau_k^+)=-\beta_n^d(\tau_k)$. Therefore, 
        \begin{equation}\label{eqn:state_der_NEP_J1}
            {x_{n,i}^d}'(\tau_k^+)=
            {x_{n,i}^d}'(\tau_k^-)+\alpha_{n}^d(\tau_k^-)\tau_{k,i}'
        \end{equation}

        \item[4.5)] Other exogenous events. Those events would not affect state derivatives, so that:
        \begin{equation}\label{eqn:state_der_NEP_other}
                {x_{n,i}^d}'(\tau_k^+)={x_{n,i}^d}'(\tau_k^-)
        \end{equation}
        
    \end{itemize}
\end{itemize}

Observe that whenever $\alpha_n^0(t)$, $n\ge2$, appears above, its value is given by 
$\alpha_{n}^0(t) = \beta_{n-1}^0(t-\frac{L_{n-1}-x_{n}^0(t)*l}{v_{n-1}})$ 
as in (\ref{eqn:flow_process_relationship}), thus capturing the interdependence of queue content derivatives between adjacent intersections $n,n-1$.
Also note that most of the queue content derivative expressions involve the event time derivative $\tau_{k,i}'$.
Therefore, to complete our analysis we need to derive these expressions through  
(\ref{eqn:event_time_derivative}) as shown next and use them in (\ref{eqn:state_der_En}), (\ref{eqn:state_der_Sn_light_switch}), (\ref{eqn:state_der_Sn_alpha}), (\ref{eqn:state_der_NEP_G2R}), (\ref{eqn:state_der_NEP_R2G}), (\ref{eqn:state_der_NEP_J0}) and (\ref{eqn:state_der_NEP_J1}).
        
\subsection{Event Time Derivatives}\label{sec:event time derivatives}
In this section, we derive the event time derivatives with respect to each of the controllable parameters $\Theta_i$, $i=1,\ldots,2N$ as formulated in (\ref{parametervector}). 

\begin{itemize}
    \item [1)] Event $E_n^d$ occurs at $\tau_k$. This is induced by [$x_n^d\downarrow 0$] so that the guard condition is $g_k = x_n^d -0=0$, which gives $\frac{\partial g_k}{\partial x_n^d}=1$, $\frac{\partial g_k}{\partial \Theta_i}=0$. And the dynamics are $f_{n,d,k-1}^x(\tau_k^-)=\alpha_n^d(\tau_k)-h_n^d$, $f_{n,d,k}^x(\tau_k^+)=0$. Then, we can derive from (\ref{eqn:event_time_derivative}) that:
    \begin{equation}\label{eqn:event_der_En}
    \tau_{k,i}'= \frac{-{x_{n,i}^d}'(\tau_k^-)}{\alpha_n^d(\tau_k)-h_n^d(\tau_k)} 
    \end{equation}
    This would complete equation (\ref{eqn:state_der_En}). And by applying it, we can get ${x_{n,i}^d}'(\tau_k^+)=0$ for state derivative case 2).

    \item[2)] Event $G2R_n^d$ occurs at $\tau_k$. This is triggered by [$z_n^d\uparrow \theta_n^d$] so that guard condition is $g_k=z_n^d-\theta_n^d$ with $\frac{\partial g_k}{\partial z_n^d}=1$ and $\frac{\partial g_k}{\partial \theta_n^d}=-1$, where $\theta_n^d$ can also be represented by $\Theta_{2n-1+d}$. We also have $f_{n,d,k}^z(\tau_k^-)= 1$. Similar to analysis in \cite{fleck_adaptive_2016}:
    \begin{equation}\label{eqn:event_der_G2R}
        \tau_{k,i}'= {\tau_{k_s,i}}'+\mathds{1}_{i=2n-1+d}
    \end{equation}
    where $k_s$ is the index of previous light switching event ($R2G_n^d$). This would complete equations (\ref{eqn:state_der_Sn_light_switch}) and (\ref{eqn:state_der_NEP_G2R}).

    \item[3)] Event $R2G_n^d$ occurs at $\tau_k$. This is triggered by [$z_n^{1-d}\uparrow \theta_n^{1-d}$] so that guard condition is $g_k=z_n^{1-d}-\theta_n^{1-d}$. Similar to analysis in last case:
    \begin{equation}\label{eqn:event_der_R2G}
        \tau_{k,i}'= {\tau_{k_s,i}}'+\mathds{1}_{i=2n-d}
    \end{equation}
    where $k_s$ is the time of previous light switching event ($G2R_n^d$). This would complete equation (\ref{eqn:state_der_NEP_R2G}).

    \item [4)]Event $J_{n-1,m}$ occurs at $\tau_k$, ($n=2,\ldots,N$, $d=0$). It is an endogenous event triggered by  $y_{n-1,m}(\tau_k)=\Delta_{n-1}(\tau_k)$, so that guard condition is $g_k=y_{n-1,m}(\tau_k)-\frac{L_{n-1}-x_{n}^d(\tau_k)*l}{v_{n-1}}$=0. Since two variables exist in the guard condition, we take derivatives of $g_k$ with respect to parameter $\Theta_i$ first. Since event time variable $\tau_k$ is directly affected by $\Theta_i$, while state variables $x_n^d(\tau_k)$ and $y_{n-1,m}(\tau_k)$ are both directly and indirectly affected by $\Theta_i$ through $\tau_k$, based on chain rule, the derivative $g_{k,i}'$ can be represented by :
    \begin{multline}\label{eqn:gk_der_y}
        g_{k,i}'=y_{n-1,m,i}'(\tau_k^-)+f_{n-1,m,k-1}^y(\tau_k^-)\tau_{k,i}' + \\ \frac{l}{v_{n-1}}({x_{n,i}^d}'(\tau_k^-)+f_{n,d,k-1}^x(\tau_k^-)\tau_{k,i}')=0
    \end{multline}
    The last event before $\tau_k$ that would cause the change of state $y$ is $G_{n-1,m}$ (start of artery flow generation). Denoting its happening time as $\tau_{k_g}$, we have $f_{n-1,m,k-2}^y(\tau_{k_g}^-)=0$ and $f_{n-1,m,k-1}^y(\tau_{k_g}^+)=1$. Also, since $y_{n-1,m}(t)=0$ right before $\tau_{k_g}$, we have $y_{n-1,m,i}'(\tau_{k_g}^-)=0$ for all $i=1,\ldots,2N$.
    So by applying (\ref{eqn:state_derivative}) and (\ref{eqn:derivative within mode}), we can get:
    \begin{equation}\label{eqn:gk_der_y_1}
        \begin{aligned}
        y_{n-1,m,i}'(\tau_k^-) &=y_{n-1,m,i}'(\tau_{k_g}^+)\\
        &=y_{n-1,m,i}'(\tau_{k_g}^-)+(0-1)\tau_{k_g,i}'\\
        &=-\tau_{k_g,i}'
        \end{aligned}
    \end{equation}
    State dynamics can be calculated from (\ref{eqn:ydynmic}) and (\ref{eqn:xdynamic}) with $\alpha_n^d(\tau_k^-)=0$:
    \begin{equation}\label{eqn:gk_der_y_2}
        f_{n-1,m,k-1}^y(\tau_k^-)=1
    \end{equation}
    \begin{equation}\label{eqn:gk_der_y_3}
        f_{n,d,k-1}^x(\tau_k^-)= -\beta_n^d(\tau_k)
    \end{equation}
    
    Then apply (\ref{eqn:gk_der_y_1}), (\ref{eqn:gk_der_y_2}) and (\ref{eqn:gk_der_y_3}) to (\ref{eqn:gk_der_y}) and rearrange:
    \begin{equation}\label{eqn:event_der_J0}
           \tau_{k,i}'=\frac{v_{n-1}}{v_{n-1}-l\beta_{n}^d(\tau_k)}(\tau_{k_g,i}'-\frac{l}{v_{n-1}}{x_{n,i}^d}'(\tau_k^-))
    \end{equation}

    where $\tau_{k_g,i}'$ is the event time derivative when $G_{n-1,m}$ occurs with [$\beta_{n-1}^0\uparrow 0$]. It can be induced by: (a)$R2G_{n-1}^0$ when $x_{n-1}^0(t)>0$, then $\tau_{k_g,i}'$ follows (\ref{eqn:event_der_R2G}); (b) $J_{n-2,\hat{m}}$ where $\hat{m}$ is the GREEN cycle index of intersection $n-2$ at $\tau_k$, then $\tau_{k_g,i}'$ follows (\ref{eqn:event_der_J0}); (c) exogenous change of $\alpha_{n-1}^0$ where $\tau_{k_g,i}'=0$.
This would complete state derivatives equations (\ref{eqn:state_der_Sn_alpha}) and (\ref{eqn:state_der_NEP_J0}).

    \item[5)]Event $J_{n-1,m}^e$ occurs at $\tau_k$ ($n=2,\ldots,N$, $d=0$). Similarly, it is an endogenous event triggered by  $r_{n-1,m}(\tau_k)=\Delta_{n-1}(\tau_k)$, with guard condition $g_k=r_{n-1,m}(\tau_k)-\frac{L_{n-1}-x_{n}^d(\tau_k)*l}{v_{n-1}}=0$. Similar to analysis in last case, the derivative of guard condition with respect to $\Theta_i$, $i=1,\ldots,2N$ is:
    \begin{multline}\label{eqn:gk_der_r}
         g_{k,i}'=r_{n-1,m,i}'(\tau_k^-)+f_{n-1,m,k-1}^r(\tau_k^-)\tau_{k,i}' + \\
         \frac{l}{v_{n-1}}({x_{n,i}^d}'(\tau_k^-)+f_{n,d,k-1}^x(\tau_k^-)\tau_{k,i}') =0
    \end{multline}
  The last event before $\tau_k$ that would cause the change of state $r$ is $G_{n-1,m}^e$ (end of artery flow generation). Denoting its happening time as $\tau_{k_e}$, we have $f_{n-1,m,k-2}^r(\tau_{k_e}^-)=0$ and $f_{n-1,m,k-1}^r(\tau_{k_e}^+)=1$. Therefore,
    \begin{equation}\label{eqn:gk_der_r_1}
    \begin{aligned}
    r_{n-1,m,i}'(\tau_k^-)&=r_{n-1,m,i}'(\tau_{k_e}^+)\\
    &=r_{n-1,m,i}'(\tau_{k_e}^-)+(0-1)\tau_{k_e,i}'\\
    &=-\tau_{k_e,i}'
    \end{aligned}
    \end{equation}
    Also by (\ref{eqn:rdynmic}) and (\ref{eqn:xdynamic}):
    \begin{equation}\label{eqn:gk_der_r_2}
        f_{n-1,m,k-1}^r(\tau_k^-)=1
    \end{equation}
    \begin{equation}\label{eqn:gk_der_r_3}
    f_{n,d,k-1}^x(\tau_k^-)=\alpha_{n}^d(\tau_k^-)-\beta_n^d(\tau_k)
    \end{equation}
    where the value of $\beta_n^d(\tau_k)$ follows (\ref{beta}).
    
    Then apply (\ref{eqn:gk_der_r_1}), (\ref{eqn:gk_der_r_2}) and (\ref{eqn:gk_der_r_3}) to (\ref{eqn:gk_der_r}) and rearrange:
\begin{multline}
\label{eqn:event_der_J1}
       \tau_{k,i}'= \frac{v_{n-1}}{v_{n-1}+(\alpha_{n}^d(\tau_k^-)-\beta_{n}^d(\tau_k))l}*\\(\tau_{k_e,i}'-\frac{l}{v_{n-1}}{x_{n,i}^d}'(\tau_k^-))
\end{multline}
where $\tau_{k_e,i}'$ is the event time derivative when $G_{n-1,m}^e$ occurs with [$\beta_{n-1}^0\downarrow 0$]. It can be induced by: (a) $G2R_{n-1}^0$ when $\alpha_{n-1}^0(\tau_k)>0$ or $x_{n-1}^0(\tau_k)>0$, then $\tau_{k_e,i}'$ follows (\ref{eqn:event_der_G2R}); (b) $J_{n-2,\hat{m}}^e$ where $\hat{m}$ is the GREEN cycle index for intersection $n-2$ at $\tau_k$, then $\tau_{k_e,i}'$ follows (\ref{eqn:event_der_J1}); (c) exogenous change of $\alpha_{n-1}^0$ where $\tau_{k_e,i}'=0$.
This would complete state derivative equation (\ref{eqn:state_der_NEP_J1}).

\end{itemize}

In summary,
\begin{equation}
\label{eqn:event_derivative_total1}
    \tau_{k,i}' =
    \begin{cases}
       \frac{-{x_{n,i}^d}'(\tau_k^-)}{\alpha_n^d(\tau_k)-h_n^d(\tau_k)}, & \text{if $E_n^d$ occurs at $\tau_k$}\\
      \tau_{k_s,i}'+\mathds{1}_{i=2n-1+d}, & \text{if $G2R_n^d$ occurs at $\tau_k$}\\ 
      \tau_{k_s,i}'+\mathds{1}_{i=2n-d}, & \text{if $R2G_n^d$ occurs at $\tau_k$}\\
    \end{cases} 
\end{equation}
\begin{equation}
\label{eqn:event_derivative_total2}
    \tau_{k,i}' =
    \begin{cases}
    (\tau_{k_g,i}'-\frac{l}{v_{n-1}}{x_{n,i}^d}'(\tau_k^-))*\\
      \quad \frac{v_{n-1}}{v_{n-1}-l\beta_{n}^d(\tau_k)}, &\text{if $J_{n-1,m}$ occurs at $\tau_k$}\\
      
      (\tau_{k_e,i}'-\frac{l}{v_{n-1}}{x_{n,i}^d}'(\tau_k^-))*\\
       \quad \frac{v_{n-1}}{v_{n-1}+(\alpha_{n}^d(\tau_k^-)-\beta_{n}^d(\tau_k))l},& \text{if $J_{n-1,m}^e$ occurs at $\tau_k$}\\
    \end{cases} 
\end{equation}
where all events were defined in Section \ref{sec:Problem Formulation} and $n=1,\ldots,N$ in (\ref{eqn:event_derivative_total1}), $n=2,\ldots,N$ in (\ref{eqn:event_derivative_total2}), while $i=1,\ldots, 2N$ in both equations.
 
\subsection{Cost Derivatives}
With all state derivatives as shown above, we can obtain the IPA cost gradient estimator as the derivative of $L(\Theta; x(0),z(0), T)$ in (\ref{eqn:L}). The IPA estimator consisting of $dL(\Theta)/d\Theta_i$, $i=1,...,2N$ is given by
\begin{equation}\label{eqn:dL}
    \frac{dL(\Theta)}{d\Theta_i} = \frac{1}{T}\sum_{n=1}^{N}\sum_{d=0}^{1}\sum_{k=1}^{K_n^d} \omega_n^d \frac{dL_{n,k}^d(\Theta)}{d\Theta_i}  
\end{equation}
where
\begin{equation}\label{eqn:Lnm}
    L_{n,k}^d(\Theta) = \int_{\xi_{n,k}^d}^{\eta_{n,k}^d}  x_n^d(\Theta,t) \,dt
\end{equation}

\begin{multline}
\label{eqn:dLnm}
    \frac{dL_{n,k}^d(\Theta)}{d\Theta_i}={x_{n,i}^d}'({\xi_{n,k}^d}^+)(t_{n,k}^{d,1}-\xi_{n,k}^d)
    +\\
    {x_{n,i}^d}'({t_{n,k}^{d,P_{n,k}^d}}^+)(\eta_{n,k}^d-t_{n,k}^{d,P_{n,k}^d})\\
    +\sum_{p=2}^{P_{n,k}^d}{x_{n,i}^d}'({t_{n,k}^{d,p}}^+)(t_{n,k}^{d,p}-t_{n,k}^{d,p-1})
\end{multline}
where $P_{n,k}^d$ is the total number of events on the observed sample path at queue $(n,d)$ within the $k$th NEP, $t_{n,k}^{d,p}$ is the observed time of $p$th event in that NEP, and $\xi_{n,k}^d$, $\eta_{n,k}^d$ are the observed occurrence times of the start and end of $k$th NEP respectively.
It is clear from (\ref{eqn:dL}) and (\ref{eqn:dLnm}) that each IPA derivative is basically the accumulation of measurable inter-event times (timers) multiplied by a corresponding state derivative. The full information needed to evaluate this IPA estimator consists of (a) event time data, which are easy to record by observing system states, and (b) state derivatives at those times, which are given by the simple iterative expression derived in \ref{sec:state derivative} and \ref{sec:event time derivatives}. Some state derivatives evaluated at an event time $\tau_k$ involve flow rates $\alpha_n^d(\tau_k)$, $h_n^d(\tau_k)$; however, these are only needed at specific events 
(e.g., at the time when $S_n^d$ is induced by a light switching event in (\ref{eqn:state_der_Sn_light_switch})), making them easy to estimate in practice, as described in the next section. Regarding the unconstrained departure rate $h_n^d(t)$ as constant also makes it easy to estimate through simple offline counting methods.
Thus, using simple online gradient-based algorithms as in (\ref{gradientopt}), we can adjust the controllable parameters to improve the overall performance (and possibly attain local optima if operating conditions do not change substantially; see also Section \ref{sec:simulation}).

\subsection{Perturbation Propagation}\label{sec:perturbation propagation}
IPA method is able to adjust the controllable parameters automatically, and when applying to multi-intersection traffic system, the adjustment of controllable parameters can synchronize the traffic lights and create certain level of ``green wave'' by propagating upstream perturbation to downstream intersections. Such propagation is induced by joining events and would affect downstream state derivatives, which captures how a perturbation in a TLC parameter $\Theta_i$ causes a perturbation in the queue content $x_{n}^0(t)$. For example, if $S_n^0$ happens induced by $J_{n-1,m}^0$ event when $u_n^0=0$, by combining (\ref{eqn:state_der_Sn_alpha}) and (\ref{eqn:event_derivative_total2}), the state derivative becomes 
\begin{multline}
    {x_{n,i}^0}'(\tau_k^+)= (\beta_n^0(\tau_k)-\alpha_n^0(\tau_k))\\
    (\tau_{k_g,i}'-\frac{l}{v_{n-1}}{x_{n,i}^0}'(\tau_k^-))
    \frac{v_{n-1}}{v_{n-1}-l\beta_n^0(\tau_k)}
\end{multline}
Since $\beta_n^0(\tau_k)=0$ and ${x_{n,i}^0}'(\tau_k^-)=0$, the equation is simplified to ${x_{n,i}^0}'(\tau_k^+)=-\alpha_n^0(\tau_k)\tau_{k_g,i}'$, where the first flow rate term can be traced to earlier upstream departure rate based on (\ref{eqn:flow_process_relationship}), and the second term of event time derivative at $G_{n-1,m}$ also depends on upstream information:(a)if it's triggered by $R2G_{n-1}^0$, $\tau_{k_g,i}'$ is an accumulation of light switching count of upstream intersection. (b) if it's triggered by $J_{n-2,\hat{m}}$ which again depends on event time derivative at $G_{n-2,\hat{m}}$. By the process of recursion, ${x_{n,i}^0}'(\tau_k^+)$ is affected by all upstream states so that any perturbation can be propagated. Similar process happens at other flow joining conditions. 
Also, observe that combining (\ref{eqn:state_der_En}) and (\ref{eqn:event_derivative_total1}) gives ${x_{n,i}^0}'(\tau_k^+)=0$ when event $E_n^0$ occurs at $\tau_k$. 
This implies that any effect on $x_{n}^0(t)$ from an upstream parameter perturbation is reset to zero after such events at intersection $n$. Therefore, for any perturbation at $n-1$ to propagate beyond $n$ requires that after a $J_{n-1,m}$ event occurs it must be followed by a $E_n^0$ before a $G2R_n^0$ event occurs so that the vehicles from $n-1$ benefiting from a positive GREEN cycle perturbation have the chance to get through $n$ without stopping; otherwise, the perturbation at $n-1$ is ``cancelled'' by $E_n^0$. 
Note that such propagation only occurs through $J_{n-1,m}$ events, which limits computation to simple derivative updates at selected events and makes this propagation analysis scalable. This also allows us to easily track how TLC parameters affect ``green waves'' along a series of intersections.

\section{Simulation Results}\label{sec:simulation}
We use Eclipse SUMO (Simulation of Urban MObility) to build a simulation environment for traffic through $N$ artery traffic light intersections, initially with $N=3$. Although IPA is independent of the arrival processes, we use Poisson processes in SUMO with corresponding rates $\bar{\alpha}$, and estimate the maximum departure rate as a constant value $h_n^d(t)=H$ for all $(n,d)$ through an offline analysis.
Since we only need the arrival flow rate at certain event times, we can estimate an instantaneous arrival rate through $\alpha_n^d(\tau_k)=N_a/t_w$, where $N_a$ denotes the number of vehicles joining queue $(n,d)$ during a time window of size $t_w$ before event time $\tau_k$; this is easy to detect and record in SUMO.  We set $H=1.3$, $v=10m/s$, and equal weights for all flows ($w_n^d=1$) throughout this section. With this setting, we have performed simulation experiments to demonstrate improvements in mean waiting times, synchronization leading to ``green waves'', adaptivity, and scalability.

\subsection{TLC for one-directional traffic}
The traffic condition is represented by Poisson arrival process with rate $\bar\alpha=[\bar\alpha_1^1,\bar\alpha_2^1,\bar\alpha_3^1, \bar\alpha_1^0]= [0.1,0.15,0.1,0.25]$, which stands for three side road directions(South-North) for each intersection and one artery direction(West-East). We set the initial controllable parameters as $\Theta_0=[(35,26),(30,20),(21,31)]$, where each tuple indicates GREEN cycle times for each intersection, consisting of artery direction and side road direction respectively. The optimal average waiting time is recorded after 20 times of parameter ($\Theta$) updates. 
The direction of each update is based on the average gradient, calculated by IPA described earlier using 10 sample paths with a length of 2000s each. Two metrics are used to show the performance: (a) Mean waiting time, and (b)Stop ratio, which is the average frequency that a vehicle stops (due to RED light effect) when driving through the artery divided by total number of intersections (3 in our case). The stop behavior can be caused by facing RED light directly, or facing GREEN light when approaching the queue tail but stuck due to ``ripple effect''. The ideal condition ``stop ratio = 0'' corresponds to a full ``green wave'', hence we our goal is to minimize the mean waiting time, and hopefully create certain level of green wave at the same time.

\begin{figure}[t]
\centering
\begin{subfigure}{.45\textwidth}\label{fig:performance_parallel_single_dir}
  \centering
  \includegraphics[width=\textwidth]{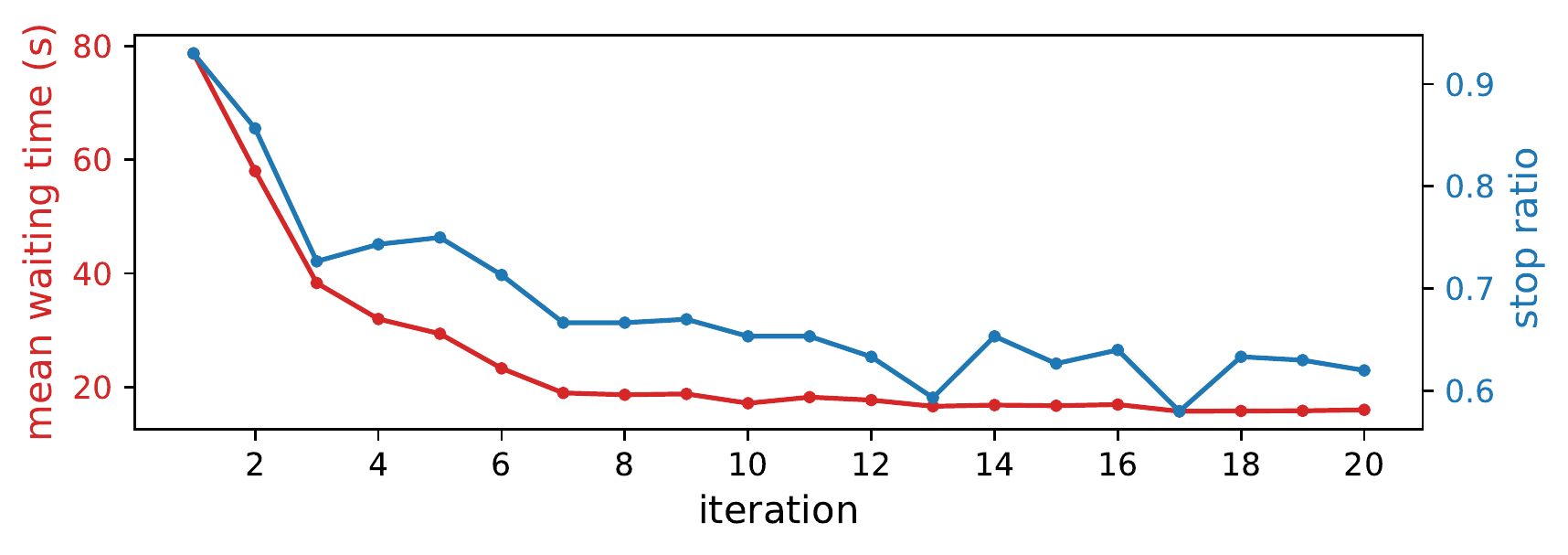} 
  \caption{Cost trajectories}
\end{subfigure}
\begin{subfigure}{.45\textwidth}\label{fig:theta_parallel_single_dir}
  \centering
  \includegraphics[width=\textwidth]{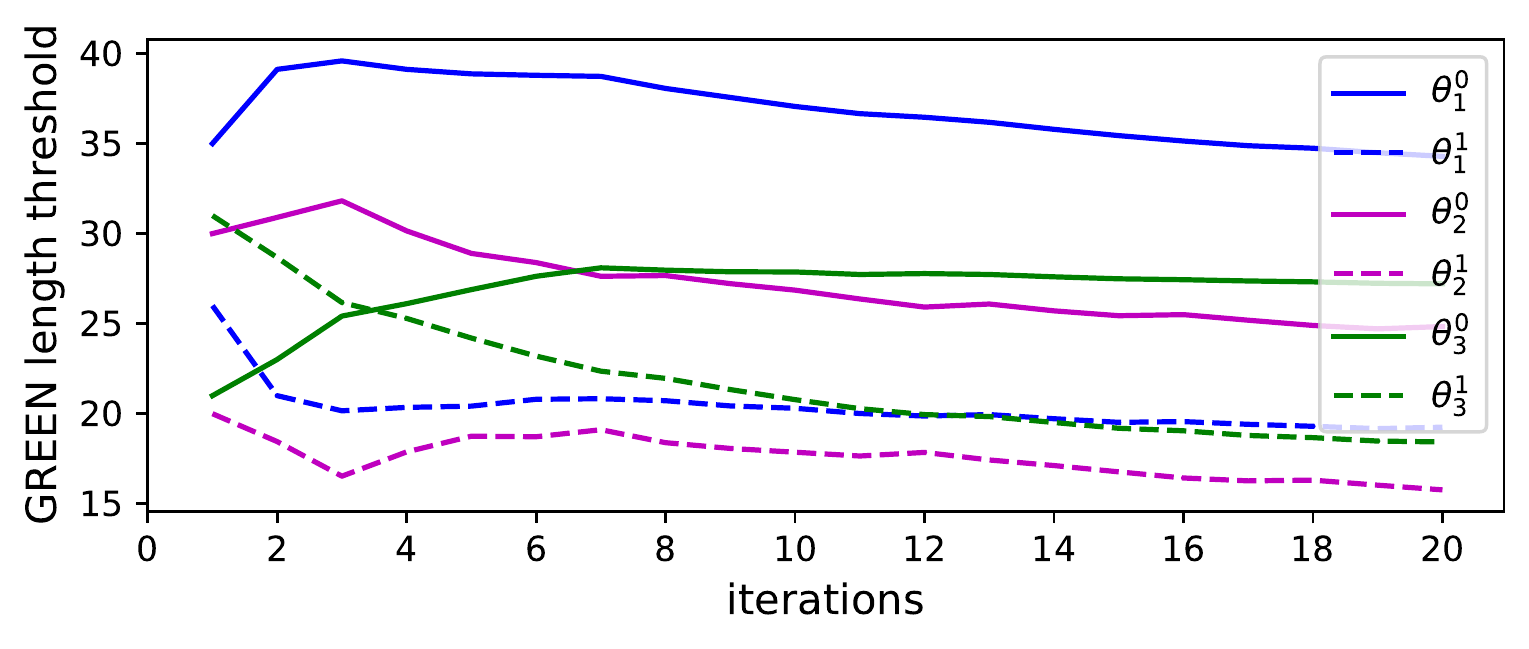}  
  \caption{Controllable parameters trajectories}
  
\end{subfigure}
\begin{subfigure}{.45\textwidth}\label{fig:performance_parallel_single_dir_sep}
  \centering
  \includegraphics[width=\textwidth]{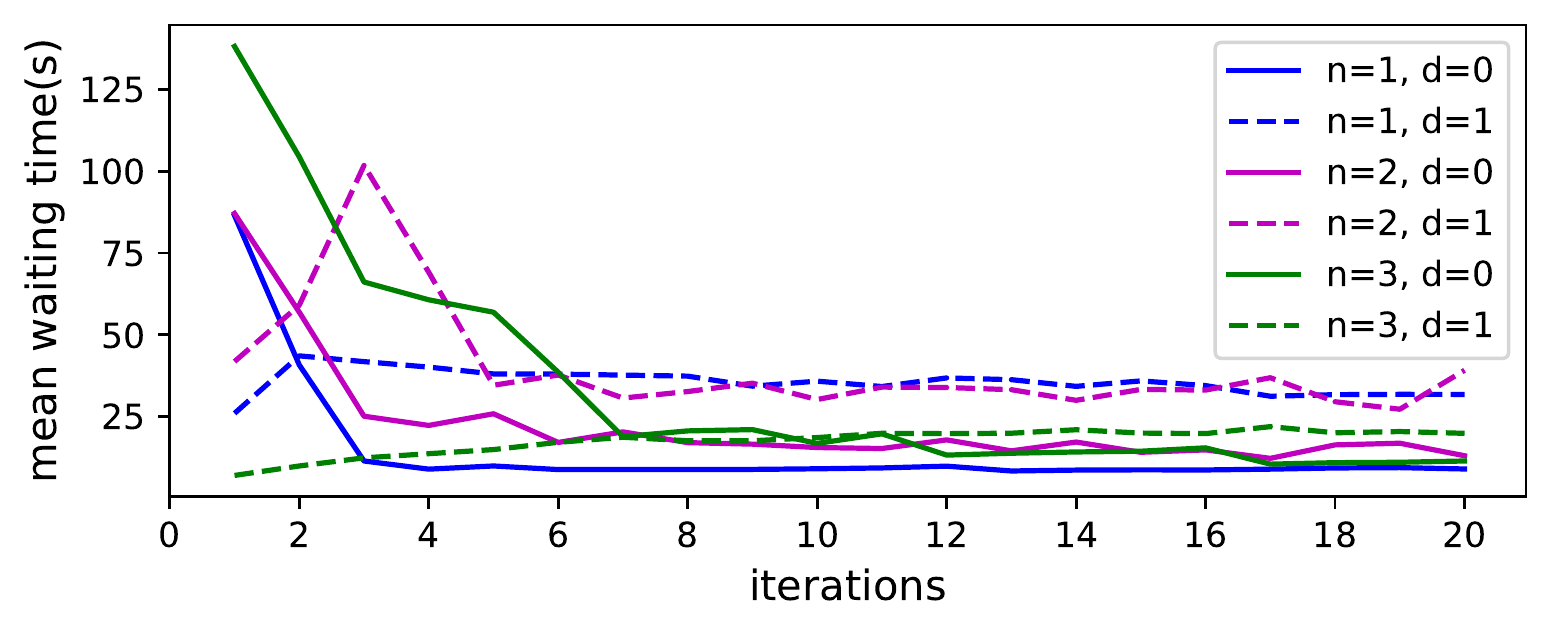}  
  \caption{Average waiting time of each direction}
\end{subfigure}
\caption{Sample cost and parameter trajectories}
\label{fig:single_dir}
\end{figure}

The resulting cost trajectories and corresponding parameter trajectories are shown in Fig.\ref{fig:single_dir}. The red trajectory shows around $75\%$ of waiting time decrease in 7 iterations compared to the value using plausible initial GREEN cycle parameters. To be more specific, by observing average waiting time trajectory of each direction, it is easy to find that the main waiting time saving is from artery direction (and the final intersection $n=3$ saves most). Mild level sacrifice of side road waiting contributes to the huge time saving for artery direction. On the other hand, we can learn from the blue stop ratio trajectory that average stop ratio also drops from 0.93 to around 0.6, which indicates 33\% less possibility that any vehicle driving through the artery would be stopped by RED light at any intersection. Combining these results, it's obvious that among 80\% of waiting time decrease, there exists optimization from traffic lights synchronization besides the part from optimization of individual GREEN cycle adjustment.





\subsection{TLC for bi-directional Traffic}
The synchronized traffic pattern provide a bandwidth of progression to achieve ``green wave'', but such bandwidth generally decreases in more constrained problem, such as two directions of an arterial or more complex geometric pattern\cite{bazzan_opportunities_2009}. In other words, two artery flows from the opposite directions would compete for the consecutive green phases, and the optimized parameter setting for single artery direction would not guarantee the good performance for traffic of opposite direction. So that in this set of simulations, we add the opposite artery traffic flow (East-West) to the same traffic setting.

\begin{figure}
    \centering
    \includegraphics[width=0.9\columnwidth]{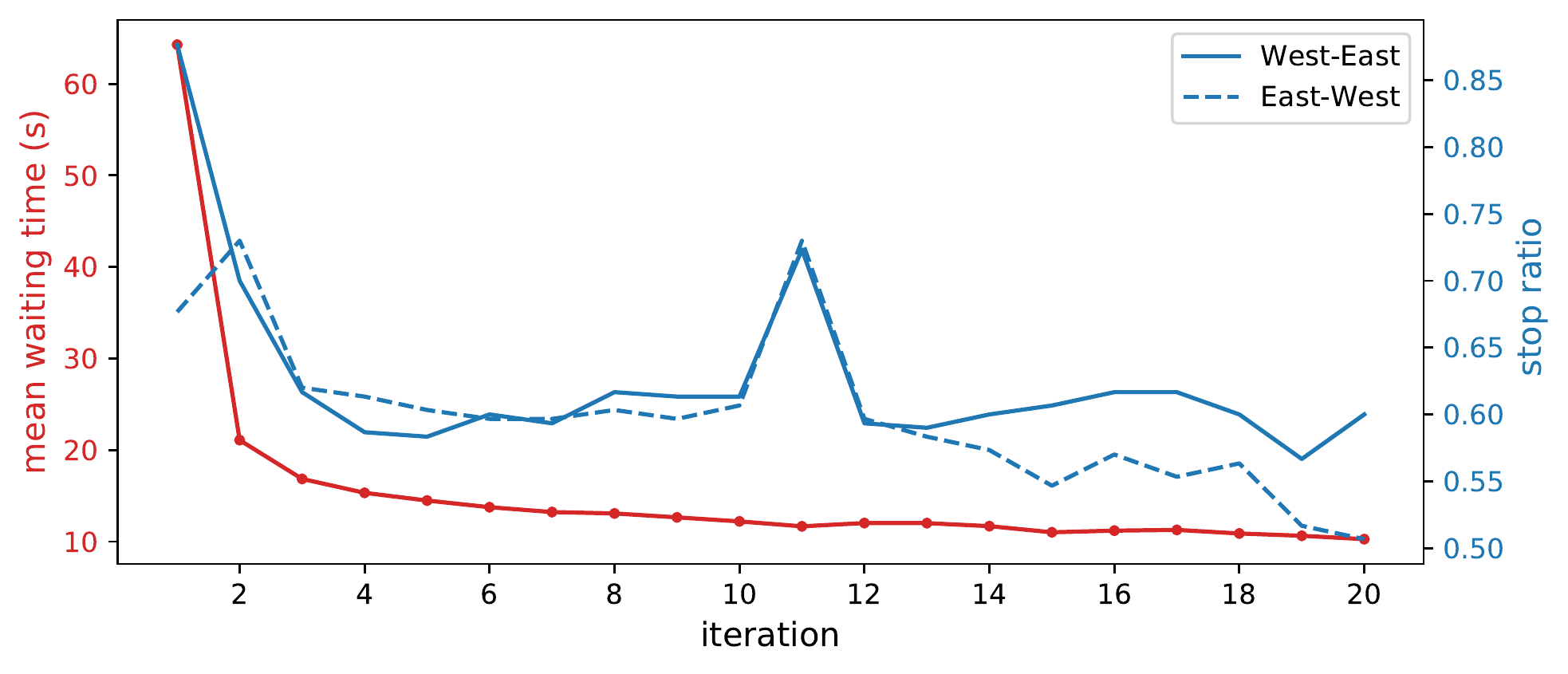}
    \caption{Sample cost trajectories of balanced bi-direction traffic}
    \label{fig:performance_parallel_bi_dir}
\end{figure}

\textbf{Balanced demands for two directions}
Similarly, we assume the vehicle arrival process to be Poisson with rate $\bar \alpha=[\bar\alpha_1^1,\bar\alpha_2^1,\bar\alpha_3^1, \bar\alpha_1^0, \bar\alpha_3^{0'}] = [0.1,0.15,0.1,0.25, 0.25]$, where the additional $\bar\alpha_3^{0'}$ stands for the Poisson rate of East-West artery flow of intersection $3$, which indicates the balanced traffic demand for two artery flows. The rest simulation configuration is the same as previous case, and the cost trajectory is presented in Fig. \ref{fig:performance_parallel_bi_dir}. Note our method can achieve good optimization result even considering bi-direction traffic flow, decreasing $80\%$ of waiting time compared to our randomly specified initial case within first 2 iterations. The stop ratio for both directions also decrease 27\% and 17\% respectively. WEST-EAST artery traffic flow shows higher ``green wave'' level than the opposite direction initially. But by synchronization effect of our method, the gap tends to be smaller with some random noise, which indicates the traffic light setting tends to be fair in terms of green wave level for traffic from opposite directions. 

\textbf{Unbalanced demands for two directions}
It is often the case that traffic demands of opposite directions are unbalanced in practice, which also induced some research about lane reversal(e.g., \cite{wollenstein-betech_planning_2021}). We then test how the unbalanced traffic demands affect the TLC optimization. With similar simulation structure to previous cases,  we set different Poisson arrival rate for EAST-WEST artery traffic demand ($\bar\alpha_3^{0'}$) and fix the rest rates.

\begin{figure}
\centering
\begin{subfigure}{.46\textwidth}
  \centering
  \includegraphics[width=\textwidth]{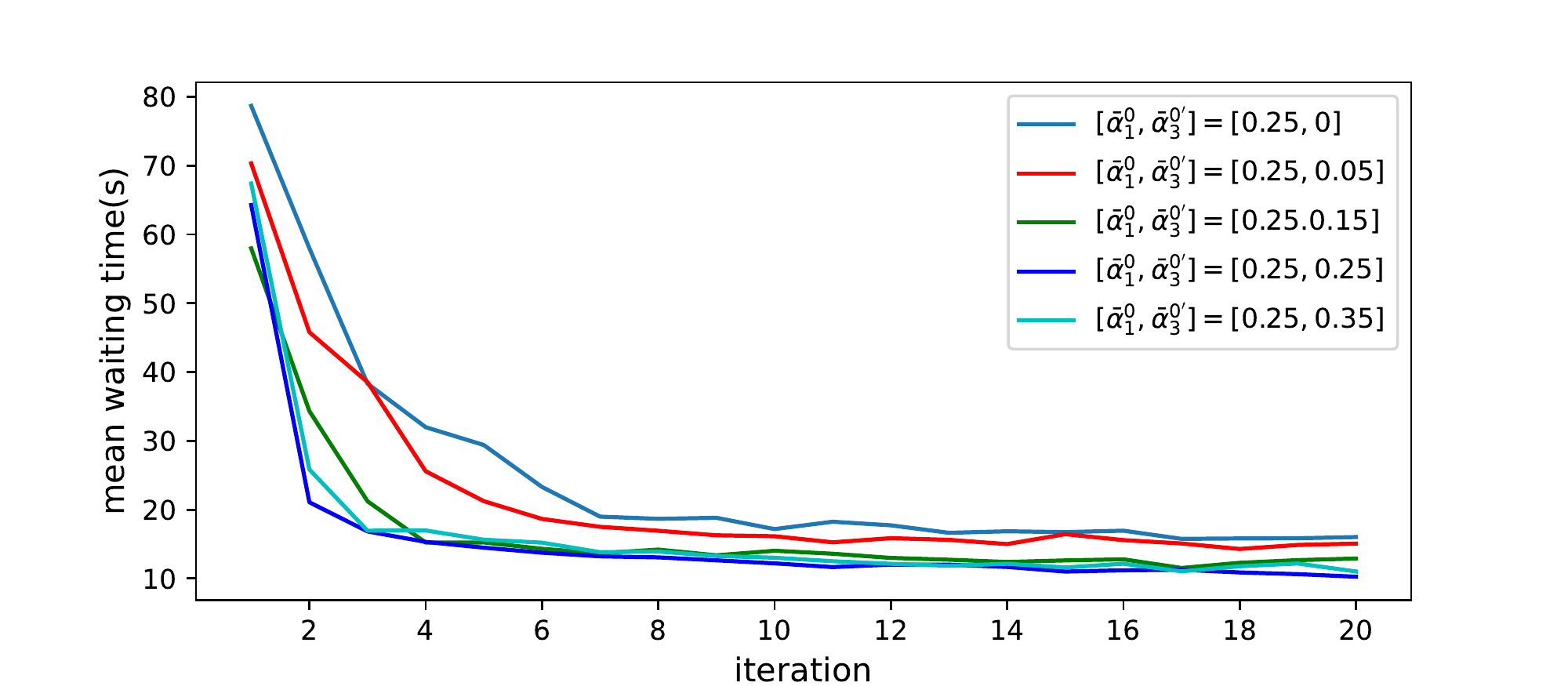} 
  \label{fig:performance_parallel_single_dir}
\end{subfigure}
\begin{subfigure}{.46\textwidth}
  \centering
  \includegraphics[width=\textwidth]{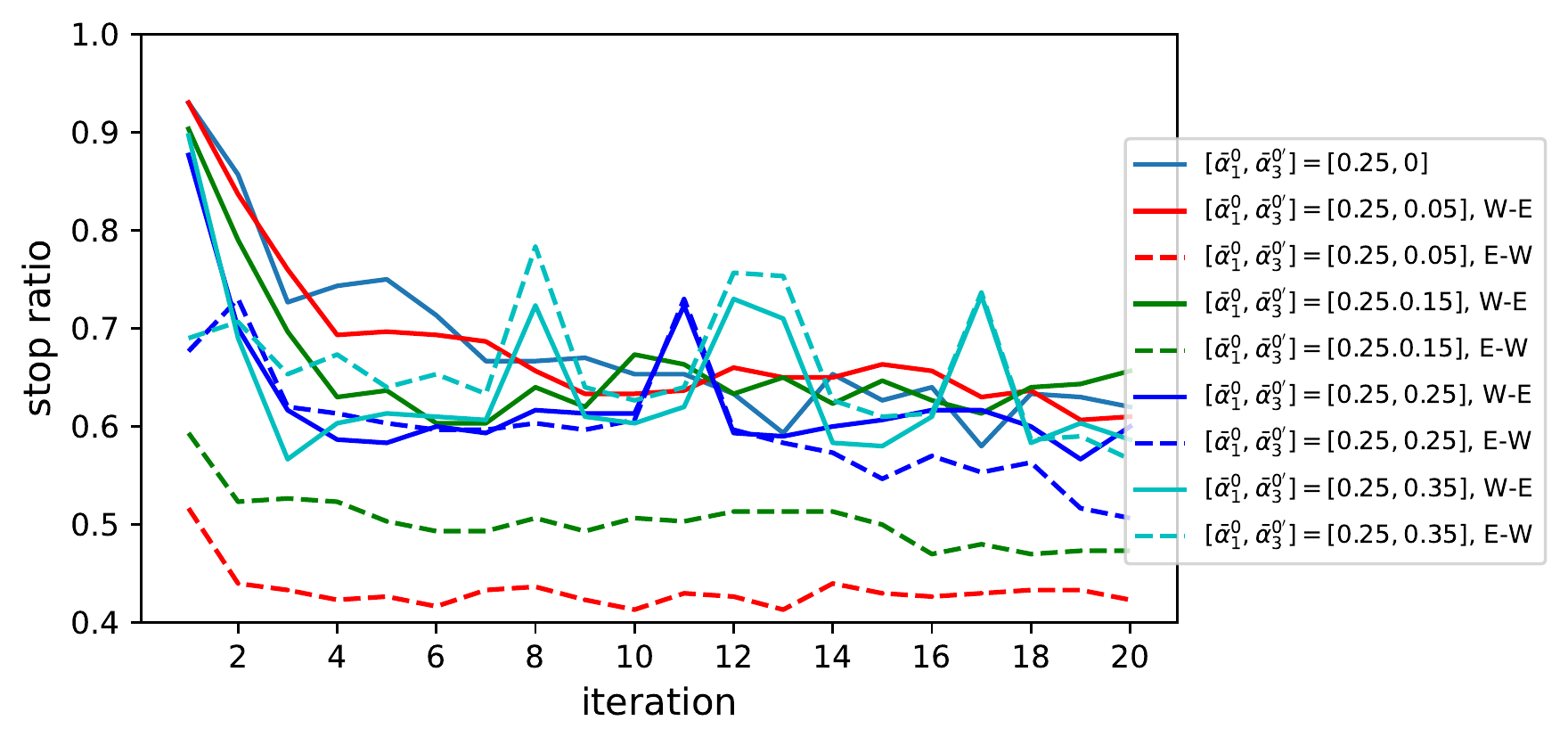}  
  \label{fig:theta_parallel_single_dir}
\end{subfigure}
\caption{Cost trajectories for different EAST-WEST traffic demands}
\label{fig:bi_dir}
\end{figure}

The specific simulation result is shown in Table \ref{table:unbalance}, and the trajectory plots are shown in Fig.\ref{fig:bi_dir}. The results show that even under extreme unbalanced traffic demand, our method has the ability to synchronize and improve the overall waiting time. Also, from stop ratio trajectory, we can roughly see that stop ratio tends to be lower for direction with lower traffic demand, which also makes sense since fewer traffic makes it less likely to build long queue, and less likely to be influenced by ``ripple effect'' and therefore easier to get through GREEN cycle without stopping.

\begin{table*}[]
\centering
\caption{Simulation results for different arterial traffic demand}
\resizebox{0.8\textwidth}{!}{%
\begin{tabular}{|l|l|l|l|l|l|}
\hline
$\bar{\alpha}$ & $J_{init}$ & $J_{opt}$ & $\Theta_{opt}$ & Cost Reduction & Stop Ratio Reduction \\ \hline
{[}0.1,0.15,0.1,0.25,0{]} & 78.69 & 16.02 & {[}(34.29,19.25),(24.84,15.77),(27.21,18.43){]} & 79.64\% & 31.0\% \\ \hline
{[}0.1,0.15,0.1,0.25,0.05{]} & 70.32 & 15.07 & {[}(33.95,19.67),(23.97,16.07),(25.61,17.76){]}  & 71.46\% & 28.2\% \\ \hline
{[}0.1,0.15,0.1,0.25,0.15{]} & 58.98 & 12.90 & {[}(33.87,18.10),(20.88,15.22),(22.36,15.88){]}   & 78.13\% & 19.9\%\\ \hline
{[}0.1,0.15,0.1,0.25,0.25{]} & 64.28 & 10.26 & {[}(31.92,18.79),(20.20,14.25),(21.85,14.36){]} & 84.04\% & 22.3\% \\ \hline
{[}0.1,0.15,0.1,0.25, 0.35{]} & 67.41 & 11.02 & {[}(32.24,19.84), (21.0,14.62),(22.55,14.79){]} & 83.65\%  & 21.1\%\\ \hline

\end{tabular}%
}
\label{table:unbalance}
\end{table*}

\subsection{Online TLC Implementation}
\begin{figure}
    \centering
    \includegraphics[width=0.9\columnwidth]{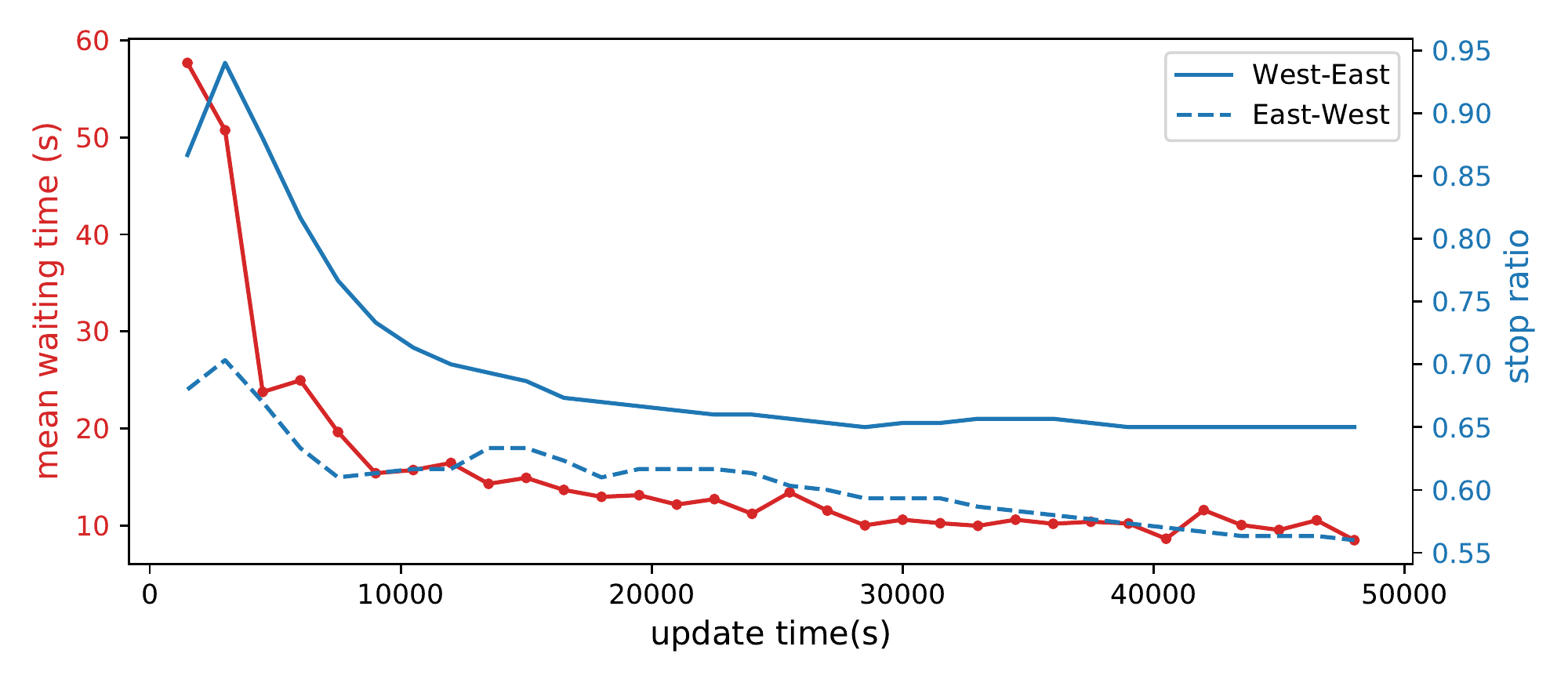}
    \caption{Sample cost trajectory for online implementation (driven by observed real-time data)}
    \label{fig:performance_seq_bi_dir}
\end{figure}

The ultimate goal of TLC is to operate on line, i.e., observe real-time traffic data and adaptively adjust the controllable parameters. 
We simulate this process by creating a single long sample path and updating IPA derivatives in (\ref{eqn:dLnm}) with every observed event occurrence (assuming some sensing capabilities for detecting events and event times).
The results are accumulated and the parameters are updated periodically. We set the initial traffic demand to be $\Bar{\alpha} = [0.1,0.15,0.1,0.25, 0.25]$, and the sample path length to be $T=43000 s$. The parameters are updated every $1500 s$ using the data collected during the most recent time window. Typical sample cost trajectories are shown in Fig.\ref{fig:performance_seq_bi_dir}. Observe that both waiting time and stop ratio for both directions significantly decrease within the first 2.5 hours and continue to improve slowly thereafter, subject to random noise due to the stochastic elements in the system. 

\subsection{TLC Adaptivity}
\begin{figure}
    \centering
    \includegraphics[width=0.9\columnwidth]{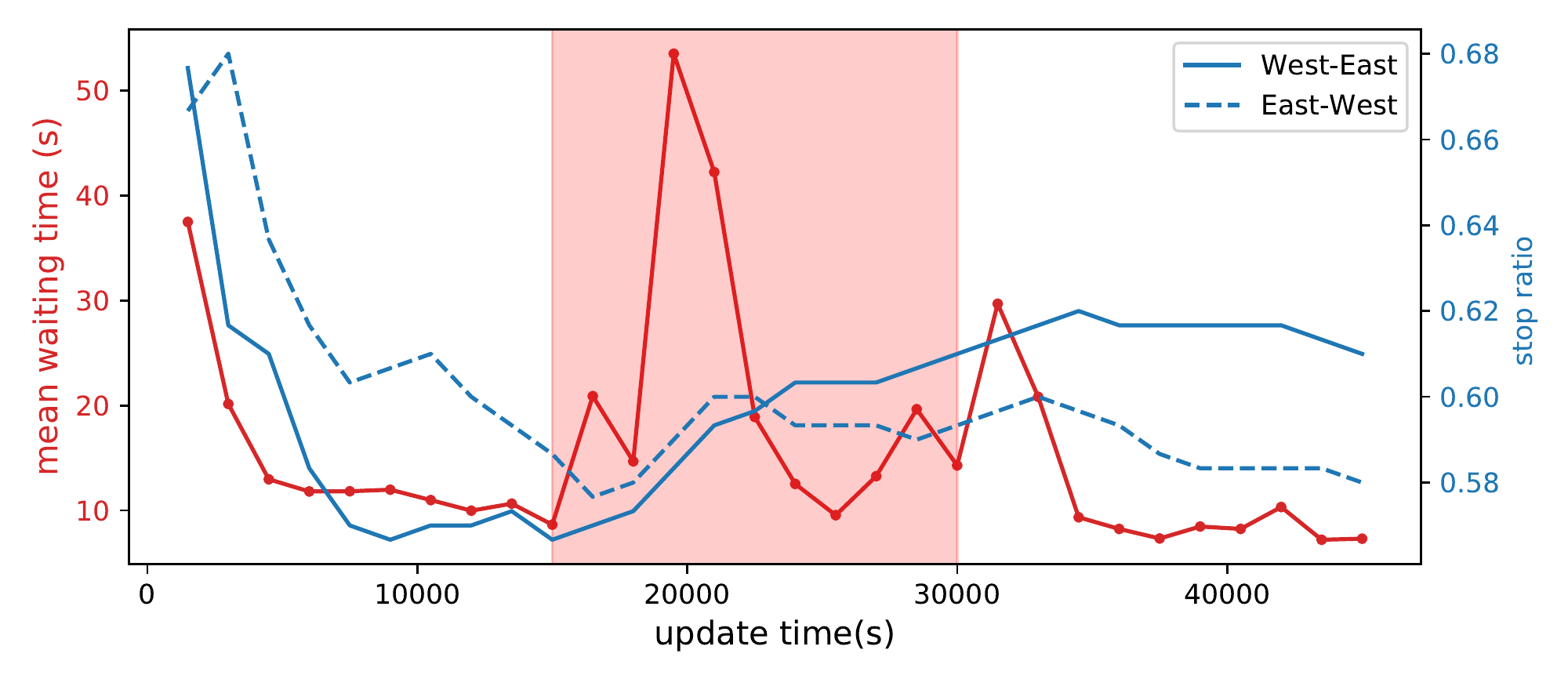}
    \caption{Sample cost trajectory for online implementation with perturbation }
    \label{fig:adaptivity_sequential}
\end{figure}

Our TLC is designed to adapt to changing traffic conditions. We illustrate this property by observing how performance changes when traffic demand is perturbed. The initial traffic demand is set by $\Bar{\alpha} = [0.1,0.1,0.1,0.2, 0.2]$, with the same initial parameters as in previous cases. We add traffic perturbations by doubling the Poisson rate
of the sideroad flow at the second intersection $t = 15000s$ and then return to the original rate at 30000s. The cost trajectory is shown in
Fig.\ref{fig:adaptivity_sequential} where the shaded area corresponds to the time interval over which traffic demand was increased. We can see that the waiting time initially decreases due to parameters adjusting. When traffic demand abruptly increases, the waiting time increases since the previously optimized parameters no longer apply to the new traffic demand. Nonetheless, they gradually adjust and converge to new optimal values after several iterations. During the adjustment period, the stop ratio for both arterial flows gradually increases, which is to be expected due to longer waiting times for a larger side road traffic demand. The stop ratios change back when the perturbation is removed.

\subsection{TLC Scalability}
\begin{figure}
    \centering
    \includegraphics[width=0.9\columnwidth]{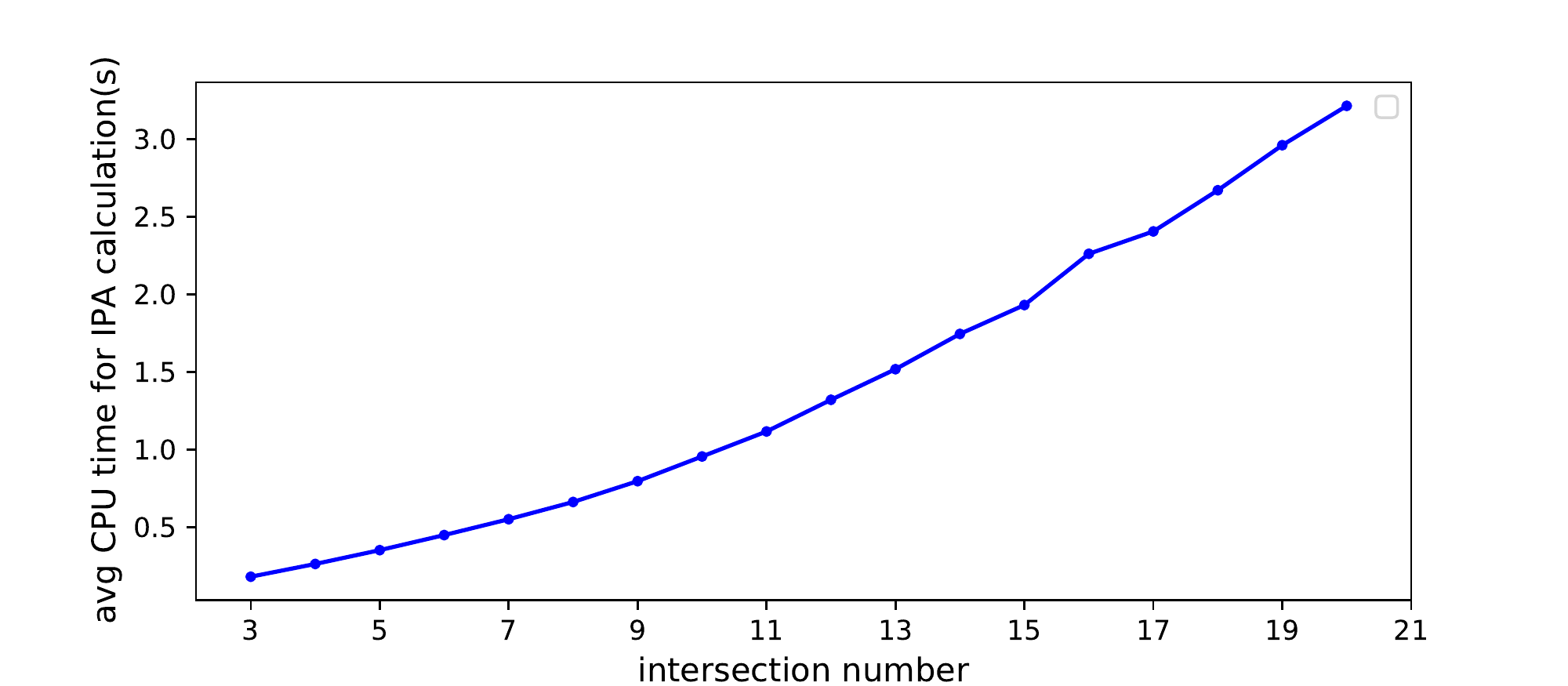}
    \caption{Scalability: CPU time as a function of the number of intersections}
    \label{fig:CPU_time}
\end{figure}

We have seen in Section \ref{section:ipa} that the IPA gradient estimation process is entirely event-driven. Therefore, the computational complexity of the TLC is linear in the number of events. This implies that our approach scales with the number of traffic lights in a network of interconnected intersections since the presence of a new intersections involves the addition of the same number of events as any other and the IPA process scales accordingly.
We illustrate this property by recording the CPU time of the IPA calculations as a function of the increasing number of intersections on the artery road as shown in Fig.\ref{fig:CPU_time}. With the number of intersections increasing for 3 to 20, the average CPU time for IPA calculations increases linearly, which motivates extending this method to large traffic networks.

\section{Conclusion and Future Work}
We have studied a TLC problem for multiple intersections in an artery with side roads, including delays for vehicle movements between intersections. We used a stochastic hybrid system model and derived IPA gradient estimators of a cost metric with respect to TLC parameters that regulate the length of GREEN cycles while taking into account the effect of transit delays by defining flow bursts and identifying flow burst joining events. Based on gradient estimates, we adjust the parameters iteratively through an online gradient-based algorithm in order to improve overall performance, with the ability to automatically adapt to changing traffic conditions. Performance is measure through the mean waiting time and a stop ratio metric which quantifies the level of ``green waves'' that can be achieved. Our next steps are to (a) add blocking events to allow for tight coupling between intersections which often occurs in practice and (b) adding more flows due to left-turn and right-turn traffic as well as bicycle and pedestrian traffic flows similar to \cite{chen2023adaptive}.

\bibliographystyle{IEEEtran}
\bibliography{references}  

\end{document}